\documentstyle[11pt,fleqn,leqno]{article}
\begin{document}

\title{Proposal for an experimental test of
the many-worlds interpretation of quantum mechanics}
\author{
        R.Plaga\thanks
        {plaga@hegra1.mppmu.mpg.de}
        \\Viktoriastr. 8
        \\80803 M\"unchen, Germany
	}
\date{}
\maketitle

\begin{abstract}
The many-worlds interpretation of quantum mechanics
predicts the formation of distinct parallel worlds
as a result of a quantum mechanical measurement.
Communication among these
parallel worlds
would experimentally rule out alternatives to
this interpretation.
A procedure for ``interworld'' exchange of information and energy,
using only state of the art quantum optical
equipment, is described.
A single ion is isolated from its environment in an ion trap.
Then a quantum mechanical measurement
with two discrete outcomes is performed on another system,
resulting in the formation of two parallel worlds.
Depending on the outcome of this measurement the ion
is excited from only one of the parallel worlds before the
ion decoheres through its interaction
with the environment. A detection of this excitation
in  the other parallel world is direct
evidence for the many-worlds interpretation.
This method could have important
practical applications in physics and beyond.
\end{abstract}
\vskip 2in
\pagebreak
\section{Introduction}
There has been a renewed
intense interest in the quantum mechanical measurement
problem recently\cite{rev}.
The reason for this is a growing dissatisfaction
with the orthodox\cite{copen} and statistical
\cite{stat} interpretations which do
not allow to derive the properties of the classical reality from
the Schr\"odinger equation even in principle.
A further problem is that both interpretations
use concepts (``reduction of the state vector'' in the former
and ``conceptual ensemble of
similarly prepared systems'' in the latter)
that are described only by words
and not mathematically, so their meaning remains vague.
Moreover
in the orthodox interpretation the human consciousness
has to play a special role in physics
(in the words of Bohr the purpose of physics is ``...
not to disclose the real essence of phenomena but only to
track down ... relations between the manifold aspects of experience''
\cite{bohr34}), a notion that does not go easy
with the majority of physicists.
\\
For simplicity
I will consider in
this paper only the simple case of
measurements with two discrete results.
A generalization to the case of
more than two outcomes is straightforward.
According to the classical book on
quantum measurements
in the orthodox interpretation by von Neumann\cite{neumann},
a quantum mechanical measurement
consists of a ``process 1'' or ``collapse of the wavefunction'':
a coherent wave function
$\psi$ (which contains a complete description of the
quantum mechanical system and of the measurement apparatus),
is suddenly converted to a statistical
mixture of $\psi_1$ which describes one possible outcome
of the measurement, and  $\psi_2$ which describes the other outcome.
This state reduction is not derived from the Schr\"odinger
equation (called ``process 2'' by von Neumann)
but introduced ad hoc to
explain the observed facts.
\\
An important progress during the last decade was the
realization that ``decoherence'' plays a decisive role
in a quantum mechanical measurement\cite{deco1}.
Decoherence explains ``process 1'' as a
loss of phase relations in the wavefunction $\psi$
of the measuring apparatus
while it interacts with the quantum system.
This loss is a continuous process and
can be quantitatively calculated in a variety
of situations\cite{deco1}
without going beyond the Schr\"odinger equation.
Process 1 needs a finite amount of
time in this view
because of its continuous nature, the so called
``decoherence time'' $\Delta$t$_{dec}$, which is very short
in most ``usual'' measurement situations
(i.e. the measurement apparatus
is macroscopic and interacts
strongly with the quantum system).
The sudden reduction envisioned by von Neumann is a very
good approximation which suffices for a description of
practical situations up to now. A complete statistical mixture
is never reached, but if one takes into account that
macroscopical
measurement apparati always interact with a large
environment, the assumption of a statistical mixture
becomes extremely good and can explain all observational facts.
\\
There remains one question (quoted here
directly from Omn\`es\cite{omnes90}): after decoherence
has taken place...``why or how does it happen
that an apparatus shows up unique and precise data
(in our case: either $\psi_1$ or $\psi_2$ is actually observed)
whereas the theory seems only to envision all possibilities
on the same footing?''.
This necessity of some mechanism in addition
to ``process 1'' (sometimes
called ``objectification''or ``actualization'') was already recognized
by von Neumann; he calls the measurement apparatus ``II''
and the apparatus ``with the actual observer'' ``III''.
He only states that the interaction between ``II'' and ``III''
``remains outside the calculation''\cite{neumann}(chapt.VI.1).
Proposals to answer Omn\`es question can be grouped
in three categories:
\\
$\bullet$ there are so called ``hidden variables'',
arising from some extension
to the Schr\"odinger equation which causes actualization
(not necessarily in a deterministic way)\cite{hv}.
A violation of the Bell inequalities
in EPR type experiments has been shown with
great precision in a variety of set ups recently\cite{eprexp}.
If one does not want to take recourse to contrived conspiracies
(see Ref.\cite{eprrec} how to exclude even these),
any hidden variable theory has to introduce
non-local interactions
as a consequence;
this would require a revision of many physical
concepts.
\\
$\bullet$ the question is declared ``meaningless'';
``actualization'' occurs without any {\it mechanism}.
e.g. Hartle states \cite{hartledis} ``We do not see it
(i.e. actualization) as
a ``problem'' for quantum mechanics.''
This standpoint is logically consistent and leads to
the so called ``logical'' \cite{rev} and ``many histories''
\cite{gellman}
interpretation of quantum mechanics. These (quite similar)
interpretations include decoherence in their description
of nature and thus go far beyond the orthodox interpretation.
Actualization is obviously crucial
for our perception of nature, but it is not
considered to be a part of
physics in this view.
Therefore these interpretations
(like the orthodox interpretation)
have to renounce the existence of an ``independent reality'',
a physical universe which exists independent of
our consciousness, Omn\`es states: ``physics is not a complete
explanation of reality...theory ceases to be identical
with reality at their ultimate encounter...''\cite{rev}.
\\
$\bullet$ a very radical and elegant answer was given
by Everett\cite{everett}:
after decoherence has taken place, the orthogonal
states
$\psi_1$ and $\psi_2$ (each also describing an independent
``split'' observer)
continue to evolve according to
the Schr\"odinger equation and have ``equal rights''.
In this view ``actualization'' is
explained as an illusion
in the brain of a human observer: after a
few decoherence times,
his weak senses and crude measuring devices are unable
to detect the increasingly weak influences of the
other ``outcome''. He therefore calls the one outcome
he can see ``the world''. The same happens with the
other outcome. For this reason DeWitt
termed the name ``many-worlds interpretation''(MWI)
for this view of nature
\cite{dewitt}. I will use the word ``universe'' to indicate
space time together with all ``worlds'' existing in it.
I call the two outcomes of a measurement
``parallel worlds'' below, because they exist in the same
Minkowskian space time.
The worlds which form as a result of a measurement
with a finite number of discrete outcomes
are usually called ``branches''.
In Hilbert space
the parallel worlds are orthogonal of course.
Together with decoherence (a concept still unknown when
Everett wrote his thesis) this idea leads to a
deterministic view of the universe
in which the human mind plays no special
role outside physics\cite{mwi2}.
\\
Section 2 contains a general discussion of the method for an experiment
to test Everett's interpretation.
Sections 3 provides a detailed analysis of a decoherence process
which is of critical importance for the experiment.
A reader mainly interested in the practical realization
of the experiment can skip this somewhat technical part
and proceed directly to section 4. Here a concrete example for
a possible technical setup is given.
In the conclusion (section 5) the predictions
of the various interpretations of quantum mechanics
for the outcome of the proposed experiment are compared, and the
potential practical importance of a result confirming the MWI
is stressed.

\section{Proposal of an experiment to
test the many-worlds interpretation}

The MWI together with decoherence
corresponds to the conceptually very simple view
that nonrelativistic quantum mechanics can be understood
by assuming only the existence of objectively real
wavefunctions whose evolution
is governed by
the usual Schr\"odinger equation,
together with the second quantization
conditions of the underlying wave field, without
any subjective or non-local elements.
It is therefore important to find experimental
tests for this interpretation.
Independent of what one thinks about
the MWI a priori, this is also a very {\it systematic} way to make
experimental progress in the question
of the interpretation of quantum mechanics, because
in the MWI the predictions for any conceivable experiment
are free from philosophical subtelties
(which can be a problem in the orthodox interpretation) or
free parameters (which often occur
in one of the many proposed hidden variable
models).
\\
I already mentioned that
decoherence only leads to approximate mixtures
(though the approximation is extremely good
in most situations)\cite{dewitt}. The
separation of worlds in the MWI is never quite complete therefore,
and there should be small influences from a parallel
world even after decoherence, {\it which must be measurable
in principle}. This has been
most clearly pointed
out by Zeh\cite{zeh72,zeh93}. In Ref.\cite{zeh72} he discusses the
possibility to observe ``probability resonances''
(later further discussed by Albrecht\cite{albrecht92}),
which occur at a singular point when the amplitudes
of $\psi_1$ and $\psi_2$ have exactly the same magnitude.
An experiment to test the MWI against the orthodox
interpretation along similar lines
was proposed by Deutsch \cite{deutsch85}.
Unfortunately it is still far from practical realization,
as it requires a computer
which remains in a quantum mechanically coherent state
during its operations and in addition possesses
artificial intelligence comparable to that of humans.
\\
I will describe an experiment for
testing the MWI with state of the art technology.
Imagine a human called Silvia
which is programmed to perform different
actions in dependence on the outcome of a quantum
mechanical measurement.
For our purposes
Silvia
might just as well be imagined e.g. as a suitably programmed
commercially available computer
connected to the experimental equipment via a CAMAC crate
instead of as a human.
As an example
Silvia sends a linearly polarized photon through
a linear polarization filter. Let the
photon be in a
state $|P\rangle$, such that the filter axis of complete transmission
is at 45$^o$ to the linear polarization plane
of the photon.
She is programmed (decides)
to switch on a microwave emitter if she will measure
that the photon passed through the
linear-polarization filter
into photomultiplier tube and to
refrain from doing so
if she will find that
the photon was absorbed by the filter.
If one assumes detectors with 100 $\%$ efficiency for simplicity,
the probablity for either outcome is 50 $\%$.
In the MWI there are two independent humans after the
measurement was performed and decoherence took place:
one which detected a photon
and switched on the emitter (called ``Silvia1'' below)
and the other
that didn't (``Silvia2'').
Could these humans (Silvia1 and 2) communicate with each other?
The standard answer in the MWI is no, because decoherence
is so complete after very short time scales
that no one of them can influence the world of the
other, which is of course necessary for communication.
\\
One could isolate a small part
of the original apparatus (before the measurement takes place)
so perfectly that it does not
immediately participate in the decoherence.
It is now possible in principle to change the state of this
isolated part {\it before} it is completely decohered
by means of an influence from only one of
the two worlds.
In this way it
could act as a ``gateway state'' between the parallel worlds.
Because it is only partially decohered, it can still be influenced by
both worlds (and in turn can influence both worlds),
thus making possible communication.
For humans an isolation on a time scale of at least seconds
would be necessary for real communication. For
the current electronic computers
a time scale of $\mu$secs and for simple macroscopic
logic electronic (e.g. in the commercial NIM standard) nsecs
would be enough to
verify the existence of the parallel world.
\\
This proposition is not
realistic if the ``gateway state'' is macroscopic, because
the required isolation would be difficult to achieve technically
(see however recent experiments with macroscopic quantum
systems e.g. Ref.\cite{maqa}).
Since the late 1970s it has become possible to perform
precision experiments on single ions stored for long times
in electromagnetic traps\cite{traps}.
I will show in section 4 that these single ions are isolated
from the environment to such a degree that
the decoherence timescale is on the order of
seconds or longer with existing technical
ion-trap equipment.
Moreover it is possible to
excite these atoms before they are
correlated with the environment to such a degree
that complete decoherence took place.
In our example above Silvia1 switches
on the microwave emitter long enough
to excite an ion in a trap with a
large probability.
After that, Silvia2 measures the state of the ion
and finds that it is excited
with some finite probability, though Silvia verified
it was in the ground state before the branching took place.
{}From that Silvia2 infers the existence of Silvia1. In an
obvious way Silvia1 and 2 can exchange informations
(bit strings of arbitrary length),
e.g. by preparing more than one isolated ion.
Single ions in traps can act as ``gateway states''
and communication between parallel worlds is possible.
\\
Let us write down the evolution of the wave function describing
the proposed experiment explicitly in several time steps.
We write the initial wave function $|{\Psi}_{t0}\rangle$
of our system (the laboratory with all its contents shortly before
the experiment begins
at time t$_0$)
as a direct product of several ``subsystems'' (in the sense
of Zurek \cite{deco1}).
The chosen factoring is somewhat arbitrary, the final results
are independent of the choice to a good approximation,
however.
\begin{equation}
|{\Psi}_{t_0}\rangle = |P\rangle \otimes
|\phi_{filter}\rangle \otimes |\phi\rangle \otimes |A\rangle
\label{t0}
\end{equation}
Here $|P\rangle$ stands for the initial state of the
photon which can be represented by the coherent superposition
$\frac{1}{\sqrt{2}}$($|P_1\rangle$+$|P_2\rangle$)
of the two polarization states of the photon (the subindex 1
indicates a polarization
plane parallel to the transmission direction of the filter,
and 2 at a 90 $^o$ angle to this direction).
$|\phi_{filter}\rangle$ describes the polarization filter,
$|\phi\rangle$ describes the laboratory
including all further experimental
equipment, possibly produced microwave
fields and Silvia. The isolated
ion in its trap is symbolized by $|A\rangle$.
A commerical linear polarization
filter is macroscopic and its Poincar\'e recurrence
time is much larger than any other time scale in the
experiment. Therefore it qualifies as ``environment'' \cite{deco1}
and some time after the photon $|P\rangle$ has interacted with the filter
(at time t$_1$) the two components of $|P\rangle$ have decohered
and we obtain to very good precision
two distinct decohered subsystems (``worlds'').
Let us call this time, when $|\phi_{filter}\rangle$ has already
decohered but the other subsystems $|\phi\rangle$ and $|A\rangle$
did not yet interact
with $|P\rangle$ ``t$_1$''
(such a time can surely be found, even if it would
be only because of the finite c).
At this time
the state of the subsystem ``photon and filter'' no
longer corresponds to any one ray in Hilbert space
(it is described by a mixture).
Rather the decoherence process has selected
two special states. While the exact nature
of these states is not yet entirely clear,
current research suggest that they are characterized
by maximal thermodynamical stability, i.e. they
are states with minimal increase in
entropy\cite{Habib}.
Let us symbolized these two orthogonal vectors in Hilbert space
in the following way:
\begin{eqnarray}
|W_1\rangle = |P_1 \phi_{filter1}\rangle
\\
|W_2\rangle = |P_2 \phi_{filter2}\rangle
\end{eqnarray}
I left out the direct product symbol $\otimes$ between the
symbols to indicate that they are in an entangled
state.These functions are very nearly orthogonal
to each other and will stay
like that forever.
However one should not conclude
that the process of decoherence is already finished.
It is finished only later when all subsystems
are decohered.
The rest of the laboratory and
the ion can still be described by pure states as can
the state of the total system at time t$_1$:
\begin{equation}
|{\Psi}_{t_1}\rangle = \frac{1}{\sqrt{2}} (|W_1\rangle + |W_2\rangle)
\otimes |\phi\rangle \otimes |A\rangle
\end{equation}
Just like the polarizer ``measured'' the two states of the
photon $|P\rangle$ via decoherence,
the subsystem ${|{\phi}\rangle}$ (including Silvia) ``measures''
the state of the polarizer. The resulting decoherence
leads to two distinct subsystems: ${|W_1\rangle}$=$|P_1 \phi_{filter1}
\phi_1\rangle$
(`photon detected world') and $|W_2\rangle$=
$|P_2 \phi_{filter2}
\phi_2\rangle$
(``no photon detected world'').
The final state at a time t$_2$ can be written as:
\begin{equation}
|{\Psi}_{t_2}\rangle = \frac{1}{\sqrt{2}} (|P_1 \phi_{filter1}
\phi_1\rangle + |P_2 \phi_{filter2}
\phi_2\rangle)
 \otimes |A\rangle
\label{t2}
\end{equation}
The ``branches''
${|W_1\rangle}$ and ${|W_2\rangle}$
are orthogonal to a very high precision,
this also guarantees the stability of the records
whether the polarized photon was detected in the further evolution
of the system.
To reach a final state at time t$_3$ in which also $|A\rangle$
is decohered
into two components (see below and section 3
for a more detailed discussion of this decoherence process), the ion
has to interact with the
rest of our system.
It is possible to excite the ion during the decoherence process,
i.e. the interaction during the time interval
$\Delta$t$_{dec}$=t$_3$-t$_2$
can excite A. When I fine tune the technical set
up
I can make sure that the time interval $\Delta$t$_{exc}$ necessary
to excite $|A\rangle$ to $|A^{\ast}\rangle$
is much smaller than $\Delta$ t$_{dec}$.
These two time scales have no direct
relation to each other.
In this case we have for the final state:
\begin{equation}
{|{\Psi}_{t3}\rangle= \frac{1}{\sqrt{2}} (|P_1 \phi_{filter1}
\phi_1 A_1^{\ast}\rangle +
|P_2 \phi_{filter2}
\phi_2 A_2^{\ast}\rangle})
\label{t3}
\end{equation}
It is of course also possible not to excite $|A\rangle$ in the course of
decoherence. The possibility of this choice allows for communication.
The excitation of an internal degree of freedom of a subsystem
does not necessarily lead to decoherence as the reader
might think at first. A counter example are
{\it Welcher Weg} detectors\cite{scully0}, in which
atoms can be
excited in micromasers without momentum transfer and
necessary loss of quantum coherence.
\\
Let us discuss in more detail
what happens when
$|A\rangle$ is excited from only one world.
Immediately after the excitation, at time
t$_2$+$\Delta$t$_{exc}$ ($\Delta$t$_{exc}$ $\ll$ $\Delta$t$_{dec}$), only
a part of the phase space in which the ion resides is excited.
It is the part corresponding to the one macroscopic
world $|W_1\rangle$ exciting the ion (macroscopic states are
very well localized in phase space\cite{tegmark1}).
After
unitary evolution of $|A\rangle$ for a short time interval of the order of
$\Delta$t$_{mix}$ = d$_{coh}$m/$\Delta$p $\simeq$ d$_{coh}${d}m/h,
the excited part of phase space begins
to overlap with the unexcited one and it is no longer possible to treat
their temporal evolution independently.
Here d$_{coh}$ is the coherence length
of the system in the branch exciting the ion,
which is extremely small for macroscopic objects\cite{tegmark1},
m is the mass of the ion and $\Delta$p is the momentum uncertainty
of a region in phase space with extension d$_{coh}$.
The momentum uncertainty $\Delta$p is approximately
given as h/{d} where d is the spatial extension of the trap.
A time scale analogous to $\Delta$t$_{mix}$ (``duration of reduction'')
in a somewhat different situation
was introduced by Dicke\cite{dicke}.
$\Delta$t$_{mix}$ can be shown
to be negligibly small for all experimental purposes
(very roughly O(10$^{-15}$ sec) for typical trap sizes ($\mu$m) and
decoherence lengths as quoted by Tegmark\cite{tegmark1}).
Because of the mentioned overlap
a measurement of the excitation of $|A\rangle$
from the other world $|W_2\rangle$, which measures another
part of phase space than a measurement from $|W_1\rangle$,
also finds the ion in an excited state.
Only after complete decoherence of $|A\rangle$
the parts of phase space seen by $|W_1\rangle$ and $|W_2\rangle$
have an independent temporal evolution.

\section{Determination of the decoherence timescale of the single ion}
I now quantitatively
calculate the time scale $\Delta$t$_{dec}$
if the decoherence of the ion wavefunction $|A\rangle$ into
$|A_{1,2}\rangle$
as defined in the previous section.
For this I will analyze the transition from eq.(\ref{t2}) to
eq.(\ref{t3}) in greater detail than before.
This analysis is independent of whether the ion is excited
between t$_2$ and t$_3$ or not.
I will use the ``dilute gas'' approximation
developed by Harris and Stodolsky \cite{stodo82,stodo89}.
The interaction of systems is treated
in terms of a series of distinct
collisions between the ion
in the trap and particles
from the rest of the system.
The correctness of this simplification
in the case of weak coupling has been verified
with a full second quantized calculation by Raffelt, Sigl and
Stodolsky\cite{raffelt}.
The
chirality states $|\pm\rangle$ of Harris and Stodolsky\cite{stodo82}
are analogous to our
macroscopic states $|W_{1,2}\rangle$ of the previous paragraph,
and their ``medium'' is the ion in the trap in our case.
Parallels between the chirality and macroscopic states
were already pointed out by Joos and Zeh\cite{joos85}.
It seems strange at first sight that a single ion
in a given ``simple''
state plays the role
of the ``medium''.
With ``simple'' I mean
that the state of the ion in its trap has only few
degrees of freedom which are
completely determined e.g. by a harmonic
oscillator wavefunctions, whereas a ``medium'' typically
has a very large number of degrees of freedom and is thus
able to
exert random influences on a system.
Take into account however that in
quantum field theory the wave field always has an infinite
number of degrees of freedom\cite{schiff}.
In the MWI it is this field which
represents all systems
and the ``simplicity''
of the state $|A\rangle$ of the ion before decoherence
at time t$_{0}$
exists {\sl only relative}
to the subsystem S$_1$=$(|P_1\rangle+|P_2\rangle) \otimes
|\phi_{filter}\rangle \otimes
|\phi\rangle$ in eq.(\ref{t0}) (Everett called the
MWI ``relative-state interpretation'' \cite{everett}).
If this subsystem decohered into two orthogonal states
$|W_{1,2}\rangle$
at time t$_2$
the ion $|A\rangle$ can no longer be in a ``simple'' state
relative to both of them, and additional degrees of freedom of the
wave function $|A\rangle$ become dynamically important.
After interaction of $|A\rangle$ with the environment,
at time t$_3$ there will
be two orthogonal components $|A_{1,2}\rangle$.
Each one has an overall centre of mass wavefunction
described e.g. by a ``simple'' harmonic oscillator
state relative to one of the worlds $|W_{1,2}\rangle$.
It is wrong to conclude from that
that they are identical, however:  $|A_{1}\rangle$ and
$|A_{2}\rangle$ are different
for the same reason that the ``copies''
produced by branching from a given macroscopic object
are not identical: their ``fine structure'' in
phase space is different.
\\
It is clear that our treatment is a
gross simplification of the
real world. An exact treatment has
been possible only for idealized
models of the environment,
e.g.: toy systems with few particles
\cite{albrecht92}, ensembles of noninteracting
harmonic oscillators\cite{legget} and scalar fields\cite{scalar}.
For the gravitational field
an exact treatment is not possible even in principle
at the moment, because we lack a quantum theory of gravity.
It has been shown experimentally though that gravitational fields
decohere if the MWI is correct\cite{page80}.
The purpose of this paper is not to improve on the treatment
of the
very difficult theoretical problem of decoherence, but to
suggest a new experimental approach on the
quantum mechanical measurement problem.
Our treatment
gives roughly the correct order of magnitude for the
decoherence time scale.
\\
Let us now define the relative states
in the sense of Everett\cite{everett} of $|A\rangle$
with respect to $|W_1\rangle$ and $|W_2\rangle$
at time t$_2$ as $|A_1\rangle$=$\frac{1}{\sqrt{2}}|A\rangle$
and $|A_2\rangle=\frac{1}{\sqrt{2}}|A\rangle$, respectively.
At time t$_2$ $|A_1\rangle$ and $|A_2\rangle$ are still the same
or ``parallel'' in Hilbert space\cite{stodo89}.
We also have
$|A\rangle$ = $\frac{1}{\sqrt{2}}(|A_1\rangle$ + $|A_2\rangle$)
a decomposition which is always possible for a pure state
according to the superposition principle.
The total wavefunction at time t$_2$ can then be written
as:
\begin{equation}
|\Psi_{t_2}\rangle=|A_1\rangle |W_1\rangle+|A_2\rangle |W_2\rangle
\end{equation}
This equation is analogous to equation (3) in Ref.\cite{stodo89}.
Further following
Harris and Stodolsky\cite{stodo82}
we now write this wavefunction in the form
of a density matrix in a basis of the Hilbert space spanned by
$|W_{1,2}\rangle$ to represent the role
of the phases in a better way:
\begin{equation}
\rho(\Psi_{t_2}) = \left( \begin{array}{cc}
\langle A_1 | A_1 \rangle & \langle A_1 | A_2 \rangle \\
\langle A_2 | A_1 \rangle & \langle A_2 | A_2 \rangle \end{array}
\right)
\label{denst2}
\end{equation}
In the initial state $\Psi_{t2}$ the ion and its environment
are uncorrelated and all elements of this matrix
have the value 1/2 in our case.
In our approximation decoherence now leads
to an exponential damping of the off-diagonal
elements of this density matrix, while the
diagonal elements remain unaffected.
At time t$_3$ the matrix is given to a very good
approximation by 1/2 the identity matrix.
The decoherence time scale in the transition
from $\Psi_{t2}$ to  $\Psi_{t3}$
is then given as the inverse of the exponential
damping time constant.
If there was no internal excitation during the process
of decoherence, $|A_1\rangle$ and $|A_2\rangle$
are identical yet distinguishable in the classical sense
(i.e. by way of their structure in phase space)
at time t$_3$.
\\
I approximate the temporal evolution
of the off-diagonal elements of $\rho$ as an
effect of repeated scatterings of particles
from ${|{W_1}\rangle}$ and ${|{W_2}\rangle}$\cite{stodo82}.
If the particles in ${|{W_{1,2}}\rangle}$
are atoms (e.g. rest-gas atoms, see below
section 4) their de Broglie wavelength
($<$ 0.1 $\AA$ at room temperature)
is much smaller than the typical spatial extension
of the wavefunction ${|{A}\rangle}$
of the ion in the trap (typically 0.1-1 $\mu$m in current
technical setups\cite{itano93}).
It is then a good approximation
for the treatment of the scattering
to assume that the initial state of the ion is approximated by
a plane wave front, and that the elastically scattered wave
of the trapped ion is approximated by
a radially outgoing wave front.
I will always use this approximation in the following
even in cases where it is less well justified
because the wavelength of the scattering
particles in ${|{W_{1,2}}\rangle}$
is equal to or larger than the spatial extension
of ${|{A}\rangle}$(e.g. for microwave
photons scattering on the ion). In this case the decoherence
time scale will be {\it larger} than my estimate
(the scattering is less ``effective'').
To demonstrate that the decoherence timescale can be
large enough to allow interworld communication,
my approach is sufficient.
Also we will see below in section 4 that
in our situation the most
effective mechanism for decoherence is
elastic scattering with rest gas atoms, for
which my assumption holds well.
\\
The diagonal element $\langle A_1{|}A_2 \rangle$ has to be multiplied
by a damping factor D for each scattering
of the ion with a particle of ${|{W_{1,2}}\rangle}$ as a target.
If $|A^{S}\rangle$ is the wavefunction of the ion
after scattering one can write:
\begin{equation}
\langle{A_1^{S}}{|}{A_2^{S}}\rangle = D \langle{A_1}{|}{A_2}\rangle
\label{damp}
\end{equation}
The damping factor after n collisions
is given as:
\begin{equation}
D_n=D^n.
\label{dampn}
\end{equation}
In the special case of elastic and isotropic scattering
and integrating over time
one has for the final state after one scattering:
\begin{eqnarray}
A_1{^S} = o( e^{ikz} + f\cdot e^{ikr}/r)
\label{in0}
\\
A_2{^S} = o( e^{ikz} + f\cdot e^{(ikr+{{\Delta}{\varphi}})}/r)
\label{in}
\end{eqnarray}
where $k$ is the wave number and $z$ the direction of relative
motion between the particle and the trapped ion.
$f$ is the scattering amplitude and $r$ the radial
distance from the ion.
$\Delta \varphi$ is a relative phase angle which takes random
values over repeated scatterings because ${|{W_{1}}\rangle}$
and ${|{W_{1}}\rangle}$ are not in phase.
The normalization factor
$o$ is given by:
\begin{equation}
o = \frac{1}{\sqrt{1+f^2/r^2}}
\label{norm}
\end{equation}
Inserting eqs.(\ref{in0},\ref{in},\ref{norm}) into eq.(\ref{damp})
and integrating over the spatial volume
one obtains:
\begin{equation}
D = o^2 = {{1}\over{{1+f^2/r^2}}} \simeq
1 - {f^2/r^2}
\end{equation}
The neglect of higher order terms
is justified in the dilute gas approximation.
For n consecutive scatterings ones gets:
\begin{equation}
D_n \simeq  (1 - {f^2/(r^2)})^n  \simeq exp(-{{f^2}n/(r^2)})
\end{equation}
Let us set f$^2$=$\sigma$/(4$\pi$), where $\sigma$
is the total elastic cross section, and n=4{$\pi$}{r$^2$}{$\phi$}t,
where $\phi$ is number of particles per unit area and time
on which the ion scatters and t the time span over which
interactions between $|A\rangle$ and $|W_{1,2}\rangle$ takes place.
The time evolution of the
diagonal elements of the ion-environment density function
is then obtained as:
\begin{equation}
D_t \simeq exp(-{\sigma}{\phi}t)
\label{dect}
\end{equation}
The decoherence time is then defined as:
\begin{equation}
\Delta t_{dec} = 1/({\sigma}{\phi})
\label{dec}
\end{equation}
This result for the decoherence time
agrees with
a different and more general calculation
by Tegmark\cite{tegmark1} for the special
case of a system that is spatially much larger
than the effective wavelength of the scattering particles.
It was exactly this case that I assumed above.
Note that Tegmark calls ``coherence time'' what I call
``decoherence time''.

\section{Practical realization of communication between parallel-worlds}

I will show that it is technically possible to
realize a system which approximates the
situation outlined in section 2. and which has
macroscopic decoherence timescales. For my discussion
I will assume the
setup which Itano et al.\cite{itano93} used
for a measurement of quantum projection noise.
This is not in order to suggest that this is
an optimal setup for inter-world communication;
I only wanted to show
that the technical capabilities to test the
MWI exist in one concrete case.
\\
Itano et al.\cite{itano93} trap single ions in radio frequency
and Penning traps. The ion (I consider $^{199}$Hg$^+$)
can be stored for hours in a vacuum of about 10$^{-9}$
atmospheres. They observe transitions
between the 6s$^2$S$_{1/2}$ F=0 and F=1 hyperfine structure
levels by applying rf fields of well-controlled frequency,
amplitude and duration. The transition is in the microwave
region (40.5 GHz). UV Lasers operated at 194 nm
are used to cool the ion, prepare its state
and to measure whether the ion is in F=0 or F=1 state
after an application of microwaves.
\\
In our example Silvia traps an ion and prepares it in the
ground state. If Silvia1 now detects a photon
after the polarization filter
she applies the rf field
resonant with the F=0 $\rightarrow$ F=1 transition,
for a time long enough to excite the ion
completely from
the ground state to the F=1 state
according to the Rabi flopping formula\cite{scully}.
According to the orthodox interpretation
she has to apply a so called ``$\pi$ pulse'' pulse of length t$_p$ and
field strength E$_{\pi}$ so that
\begin{equation}
E_{\pi}=(\pi{\hbar})/(t_p{\wp})
\label{pi}
\end{equation}
$\wp$ is the magnetic dipole transition element
between the F=0 and F=1 states (the transition is
forbidden for electric dipole radiation) which is given in
good aproximation by the Bohr magneton because the
wavefunctions of the two states are quite similar.
Let us assume that Silvia1 applies a pulse which is
a factor $\sqrt{2}$ longer to compensate for
the fact that Silvia2 does not apply any pulse
(``MWI $\pi$ pulse'').
\\
This whole action
will take something like a second at least (for
a mechanical ``Silvia'' it could be performed faster,
certainly within a $\mu$sec).
Silvia2 waits for
a certain time to allow Silvia1 to apply the microwave field.
After this she applies a Laser field to determine
the state of the ion. If the MWI interpretation is
correct, Silvia2 will find it in a fraction p of the
experiments
in the F=1 case prepared by Silvia1.
If the inelastic microwave excitation
is the only interaction of ion with the environment
(i.e. the ion is completely isolated)
we get for the damping factor due to excitation
according to eq.(\ref{dect}):
\begin{equation}
D_t \simeq exp(-{\sigma_{exc}}{\phi}t)
\label{indec}
\end{equation}
here ${\sigma_{exc}}$ is the cross section
of the ion for excitation from the F=0 to the F=1
level with resonant microwave radiation.
t is the time period for which the rf field was applied, and
$\phi$ is the flux of the exciting radiation.
The excitation probability is given as:
\begin{equation}
p=sin^2({\nu}t)
\end{equation}
where $\nu$=$\wp$E$_{\pi}$/($\hbar$2$\sqrt{2}$).
For a ``MWI $\pi$ pulse'' p is 1
and D$_t$ can be easily evaluated as
1/e. Intuitively one can say, that in this
situation only one full interaction took place (the absorption
of one microwave photon).
Complete decoherence needs
more than one interaction so D$_n$ is much larger than zero.
Normally Silvia2 will completely decohere
the ion when determining its state with
the method decribed by Itano et al.\cite{itano93},
because the detection of the fluorescence radiation is
very inefficient, and many inelastic collisions
of 194 nm photons take place for a state determination.
\\
This calculation is only correct
in the \mbox{``one-and-only-one interaction''} approximation of
Stodolsky\cite{stodo89}
in which the different collisions of the ion on other particles
are treated as completely independent.
It is unavoidable
in our situation that there is ``feedback'',
i.e. a given collision acts on a wavefunction
of the ion which is already decohered to some degree by
the previous collisions.
As a result
the excitation of $|A_2\rangle$ will be less effective and p will
be somewhat smaller than 1.
Its exact value depends on the detailed geometry
of the experimental setup but is clearly never much smaller than
1, because in the absence of other mechanisms
the correlation has its origin in the
inelastic scattering of the ion.
I find with a numerical calculation that e.g. a ``MWI $\pi$ pulse''
applied in world 1 would lead to p=0.163 in the ``feedback''
case, versus p=1.0 in the
``one-and-only-one interaction'' approximation.
In this calculation I made the simplifying assumption
that D develops strictly according to eq.(\ref{dect}).
Itano et al.\cite{itano93}
repeated the cycle ``preparation-rf field application-measurement''
for hundreds of times in their experiment so also
values of p much smaller than 1 would be measurable.
\\
We have to check if decoherence by other
sources can be avoided for at least a few seconds
so that the assumption of complete isolation of the ion made
in the previous paragraph is justified. These sources are:
\\
a. scattering of remnant gas atoms
in the trap on the ions
\\
b. elastic scattering of the microwave field on the ion
\\
c. interaction with the constraining fields holding the ion
\\
Note that only b. is in principle unavoidable, the others
could be avoided with a more advanced technology.
For contribution a. I get, inserting typical operating
parameters of the set up used by
Itano et al.\cite{itano93} into eq.(\ref{dec}):
\begin{equation}
\Delta t_{dec} = 8 \left({{2.4{\cdot}10{^{-18}m^2}\over{\sigma_c}}}\right)
\left({{T}\over{300K^o}}\right) \left({{nbar}\over{p}}\right) sec
\end{equation}
here $\sigma_c$ is the elastic cross section;
its size (for room temperature) has been taken
for H$_2$-Hg collisions (at room temperature)
from the calculation of
Bernstein \cite{bernstein}. T is the temperature,
its dependence here does not take into account the
change of $\sigma_c$ with energy (which is however
very small around room temperature). p is the rest-gas pressure.
It is possible
to achieve vacua much better than a nbar
in ion traps (see e.g.Ref. \cite{diedrich}).
\\
For b. one gets in the same way the
decoherence time scale of elastic scattering
of a microwave field with a frequency $\omega$
and an intensity that
effects a $\pi$ pulse in t$_p$ seconds.
For the flux $\phi$ in eq.(\ref{dec}) I set:
\begin{equation}
\phi = {{{\epsilon_0} c E_{\pi}^2}\over{\hbar {\omega}}}
\end{equation}
where E$_{\pi}$ is the electric field strength
of a MWI-$\pi$ pulse(eq.(\ref{pi}).
Inserting this relation gives:
\begin{equation}
\Delta t_{dec} \simeq 2.8 \cdot 10^{22} \left({{\wp}\over{{\mu_B}/c}}\right)
\left({{t_p}\over{sec}}\right) \left({{5.2{\cdot}10^{-40}m^2}
\over{\sigma}}\right) \left({{\omega}\over{40.5 Ghz}}\right) sec
\label{muw}
\end{equation}
The cross section is the Thompson cross section which
I averaged over scattering angle. The Rayleigh
cross section is negligible in our situation.
\\
Case c. is treated in a similar way because it is
well known that only time dependent fields can cause
decoherence \cite{deco1}. Even for Penning
traps with static fields it is impossible
to prevent residual time variability with a
fraction f$_v$ of the total field strength.
Without load (as in our case) f$_v$ $\simeq$
10$^{-10}$ is achievable for static confining fields E$_c$
with a strength of about 1000 V/m typical for the traps
used by Itano et al.
(their ion-trap setup is described in Gilbert et al.\cite{itano88}).
The ``worst'' case (leading to
the shortest decoherence time) is a variability $\omega$ on a time
scale similar to the duration of the experiment.
For this case one then obtains:
\begin{equation}
\Delta t_{dec} \simeq 76
 \left({{5.2{\cdot}10^{-40}m^2}
\over{\sigma}}\right) \left({{\omega}\over{1 Hz}}\right)
 \left({{1000 V/m}\over{E_c}}\right)^2 f_v^{-2} sec
\label{field}
\end{equation}
Though it is not of critical importance
for our problem,
it is easy to show that the decoherence time scale induced
by UV Lasers used by Itano et al.\cite{itano93}
via Rayleigh scattering is on a time scale
of many years.
This surprising ineffectiveness of light to decohere
wave functions was already noticed by Joos and Zeh in the connection
with chiral eigenstates of molecules \cite{joos85}.
As pointed out in the previous paragraph
eqs.(\ref{muw},\ref{field}) are expected to underestimate
the true decoherence time, because I assumed
in their derivation that the
wavelength of the particles on which the ion
scatters is much smaller than the spatial extension
of the ion wavefunction, which does not hold in typical setups.
\\
The reader might object that something has to
be wrong with my proposal
because it violates energy conservation
in a given world
(Silvia2 could receive energy from a parallel
world). Fundamental principles (like invariance to time translations
\cite{noether})
require energy conservation only for the whole universe however,
and not for single branches which are very special
entities singled out by individual humans.
Because the energy Silvia2 receives is always lost by Silvia1
there is no violation of energy conservation in the universe.
Dicke found some time ago
that energy conservation is violated
in certain quantum mechanical measurement setups
for arbitrarily long times\cite{dicke}.
He holds that this poses no serious problem because the
expectation value for the amount of energy violation
turns out to be zero (i.e. repeating the measurement
many times, energy is lost as often as it is gained).
In the conventional interpretation of quantum mechanics
there seems to be a problem however, because Dicke's result means
that e.g. the fundamental principle of time tanslation
invariance would be violated on macroscopic time
scales. In the
MWI Dicke's situation corresponds to
worlds which have a different energy expectation value of the
system immediately after they were
created due to branching (one is higher and the other lower
than the one before branching\cite{dicke}).
The average of their
energy expectation values is the energy expectation value
before the
branching, and energy conservation holds
at all times.
This ``restoration of conservation laws'' in the MWI, which arises
when all branches of the quantum state are considered together
was already pointed out by Elitzur and Vaidman\cite{eli}.

\section{Conclusion}

The prediction of the orthodox interpretation \cite{neumann}
is that the ion in our example experiment is never observed in
an excited state by Silvia2: the measurement is
surely finished
after the photon from the polarization filter has
not been detected
by Silvia2 and thereafter only Silvia2 exists.
The ``logical'' and ``many histories''
interpretations \cite{gellman}
probably lead to a similar expectation, though it
is not completely clear to me what their quantitative prediction
would be. Hidden variable
models are devised in order to ``destroy'' Silvia1;
their predicition is therefore the same
as in the orthodox interpretation by definition.
For the MWI
it has been shown
in the previous sections
that inter-world communication
on a time scale of minutes should be possible with
state of the art quantum-optical equipment.
The experimental verification of this possibility
would thus rule out the above mentioned alternatives
to the MWI.
\\
The limiting factor in extending $\Delta$ t$_{dec}$ even further
(i.e. in ``keeping the communication channel open for longer'')
seems to be the rest gas in the vacuum
of the ion trap at the moment.
The fascinating problem of how to optimize
the communication in order to transfer large amounts
of data (e.g. TV pictures)
would be beyond the scope of this paper.
\\
The detection of parallel worlds
would finally clarify the fundamentals of
nonrelativistic quantum mechanics:
nature would have an objective
deterministic reality completely independent
of human consciousness and
fully described by the Schr\"odinger equation
together with the second quantization conditions
for the wave field.
To communicate with parallel worlds goes of
course completely against ``common sense'', but
it does not lead to any inconsistencies or violations
of known physical principles.
A similar opinion was voiced by Polchinski\cite{polchinski}
who showed that interworld communication is
possible within Weinberg's nonlinear quantum mechanics.
The recent speculation of Gell-Mann and Hartle\cite{goblin}
about a possible
communication with ``goblin worlds'' has also
certain parallels with the proposal of this paper.
\\
The applications of this effect in physics would be manifold
e.g. in the investigation of Chaos or for improving
statistics in the study of rare processes.
Outside physics inter-world communication
would lead to truly mind-boggling possibilities, e.g. in
psychological research or for the extension of
computing capabilities in computers and humans.

{\LARGE Acknowledgements}
\\
I thank M.Jarnot, R.Mirzoyan and
S.Pezzoni for discussions and
helpful remarks on the manuscript
and L.Stodolsky for explanations about the nature of decoherence.
My sister Fritzi Plaga encouraged me in an important
way to work on the subject of the present paper.
\end{document}